\documentstyle[aps,preprint]{revtex}
\begin{document}

\begin{center}
{\huge On the Momentum Distribution and \\[5mm]
Condensate Fraction in the Bose Liquid}\\ [5mm]
{\large A. Yu. Cherny
\footnote{Permanent address: Obninsk Institute of Nuclear
Power Engineering,
Studgorodok, Obninsk, 249020, Russia}
and A.A. Shanenko} \\[3mm]
{\it Bogoliubov Laboratory of Theoretical Physics \\
Joint Institute for Nuclear Research \\
141980, Dubna, Moscow region, Russia}\\

\end{center}

\vspace{0.3cm}

\begin{abstract}
The model recently proposed in Ref.1 is used to derive linear
integro--differential equations whose solutions provide
reasonable estimates for the momentum distribution and
condensate fraction in interacting many--boson system at zero
temperature. An advantage of these equations is that they can
be employed in the weak coupling regime and beyond. As an
example, analytical treatment of the weak coupling case is
given.
\end{abstract}
\vspace{0.5cm}
\noindent
{\bf PACS: 05.30.Jp, 67.90.+z}
\vspace{1cm}

A new way of investigating spatial particle correlations
has recently been proposed for the many--boson system in the ground
state~\onlinecite{shan1}. It results in integro--differential
equations for the pair distribution function. They accurately take
into account the short--range boson correlations and, in the weak
coupling regime, yield thermodynamics which reasonably agrees with
the data of the Bogoliubov approach~\onlinecite{shan1,shan2}. The
structure of the derived equations is similar to that of the
Born--Green equation~\onlinecite{born} for the pair distribution
function of classical liquids. This gives an optimistic view on
the possibility of using methods developed for classical liquid
in investigating the Bose one. The most important peculiarity
of the new approach is that it does not assume a small depletion
of the zero momentum state. Moreover, the density of the condensate
particles is not at all a parameter contained by the
integro--differential equations. So, a question arises how to
calculate the condensate density within the model of Ref.1. In
addition, it is useful to clarify how to determine the momentum
distribution of particles in the Bose liquid with the help of the
integro--differential equations mentioned above. Answering these
questions is the aim of the present Letter.

The phenomenon of the Bose--Einstein condensation consists in the
macroscopic occupation of the zero momentum state so that for the
total density of bosons we have (in the thermodynamic limit)
\begin{equation}
n=n_0+\frac{1}{(2\pi)^3}\;\int\,n(q)\,d^3q\; ,
\label{1}
\end{equation}
where $n_0$ denotes the density of condensate particles, $n(q)$
stands for the distribution of bosons over nonzero momenta.
According to expression~(\ref{1}), $n_0$ can be found with the
known distribution $n(q)$ at any given $n$. Hence, we would be able
to calculate $n_0$ within the approach of Ref.1 if we managed to
derive an expression for $n(q)$ in terms directly connected with the
spatial correlations. An attempt is natural to get this expression
with the help of the ground--state energy $E$ because there are two
important relations linking it with both the pair distribution
function and $n(q)$. These relations are the consequences of the
well--known statement which is often called the Hellmann--Feynman
theorem and results in the following equality:
\begin{equation}
\delta E=\langle \psi\mid \delta H \mid \psi\rangle \;,
\label{2}
\end{equation}
where $\delta E$ and $\delta H$ are infinitesimal changes of the
ground--state energy and Hamiltonian, respectively, and $\psi$
denotes the ground--state wave function. The first relation is given
by~\onlinecite{pines}
\begin{equation}
E=E_{id}+\frac{N^2}{2V}\;\int\limits_{0}^{1}\,d\gamma\;
\int g(r;\gamma,n)\,\Phi(r)\,d^3r\;.
\label{3}
\end{equation}
Here $E_{id}$ stands for the energy of noninteracting particles
($E_{id}=0$ represents the most general case for bosons), $V$ is the
system volume and $N=n\,V\;.$ Besides, $\gamma$ denotes
the coupling constant, $\Phi(r)$ stands for the interparticle
potential and $g(r;\gamma,n)$ is the pair distribution function.
Note that below, for the sake of brevity, the notation $g(r)$ is
also used instead of $g(r;\gamma,n)$ so that $g(r) \equiv
g(r;\gamma,n)\;.$ The quantities related to $g(r)$ are handled in
the same way. The second important relation mentioned above concerns
the functional derivative of $E$ with respect to the one--particle
kinetic energy $\displaystyle T(q)=\frac{\hbar^2 q^2}{2m}$ and is
expressed as
\begin{equation}
\frac{\delta E}{\delta T(q)}=\frac{V}{(2\pi)^3}\;n(q), \quad q
                                                     \not= 0\;.
\label{4}
\end{equation}
Combining (\ref{3}) and (\ref{4}) leads to the following equality:
\begin{equation}
n(q)=4\,\pi^3\,n^2\;\int\limits_{0}^{1}\,d\gamma\,\int \,\Phi(r)\;
\frac{\delta g(r;\gamma,n)}{\delta T(q)}\;d^3r\;.
\label{5}
\end{equation}

Now, we have got the connection of $n(q)$ with $g(r)$. However, to
employ it one needs to know more than simply the dependence of the
pair boson distribution on $r$, $\gamma$ and $n\;.$ One should also
have the knowledge of the functional dependence of $g(r)$ on $T(q)
\,.$ The integro--differential equations for $g(r)$ proposed in
Ref.1, turn out to provide all the necessary information. Let us
consider the simplest of them, to shorten the further reasoning
without loss of generality. This equation may be written in the
following form:
\begin{equation}
\frac{\hbar^2\, \nabla^2\,u(r)}{m\,\Bigl(1+u(r)\Bigr)}=
\gamma\,\Phi(r)+n\,\gamma \; \int\, \Phi(\mid{\bf r} - {\bf y} \mid)\;
\left(u^2(y)+2u\,(y)\right)\,d^3y\;,
\label{6}
\end{equation}
which is convenient in reaching our aim. Here $\displaystyle
u(r)=g^{1/2}(r)-1\;.$ Taking the functional derivatives of the left--
and right--hand sides (l.h.s. and r.h.s.) of (\ref{6}) and, then,
equating one to another, we are able to find an equation for
$$
K_{{\bf q}}({\bf r}) \equiv \frac{\delta u(r)}{\delta T(q)}\;,
$$
and $K_{{\bf q}}({\bf r})$ may be used in evaluating $n(q)$ with
the obvious relation
\begin{equation}
\frac{\delta g(r)}{\delta T(q)} = 2\,\biggl(1+u(r)\biggr)\,
            K_{{\bf q}}({\bf r})\;.
\label{7}
\end{equation}
While calculating, one should realize that the form of the operator
$\displaystyle \frac{\hbar^2}{m} \nabla^2$ is completely specified
by the shape of $T(q)\;.$ Hence, the functional derivative and
operator $\displaystyle \frac{\hbar^2}{m}\nabla^2$ are not
commutative in the situation considered. Now, let us perturb the one
particle kinetic energy replacing $T(q)$ by $T(q)+\delta T(q)\;.$
Working only to the first order in the perturbation,
for the change of the l.h.s. of (\ref{6}) we have
\begin{equation}
\delta (l.h.s.)=-\,{2 \over 1+u}\,\delta
          \left(-\frac{\hbar^2}{2m}\;\nabla^2 u \right)
              - {\hbar^2 \over m\,(1+u)^2}\;\delta u\,\nabla^2 u.
\label{8}
\end{equation}
The relation
$$
-\frac{\hbar^2}{2m}\nabla^2 u(r)=\frac{1}{(2\pi)^3}\;
   \int\,T(q)\;\widetilde u(q) \exp(i {\bf q}\,{\bf r})\;d^3q\;,
$$
where $\widetilde u(q)$ stands for the Fourier transform of $u(r)$,
makes it possible to get
\begin{equation}
\delta\left(-\frac{\hbar^2}{2m}\nabla^2 u\right)= -
   \frac{\hbar^2}{2m}\nabla^2\left(\delta u\right)+
      \frac{1}{(2\pi)^3}\;\int\;\delta T(q)\;\widetilde u(q)
         \exp(i {\bf q}{\bf r})\;d^3q\;.
\label{9}
\end{equation}
Inserting (\ref{9}) into (\ref{8}) we arrive at
\begin{eqnarray}
\delta (l.h.s.)&=&{\hbar^2\over m\,(1+u)}\,
                         \nabla^2\left(\delta u\right)
	  -{\hbar^2\over m\,(1+u)^2}\,\delta u\;\nabla^2 u
                                           \;-\nonumber\\[5mm]
&&-\;\frac{2}{1+u}\;\int\;\delta T(q)\;\widetilde u(q)
          \;\frac{\displaystyle \exp(i
                {\bf q}{\bf r})}{(2\pi)^3}\,\;d^3q\;.
\label{10}
\end{eqnarray}
In its turn, the leading term produced by the perturbation in the
r.h.s. of (\ref{6}) is given by
\begin{equation}
\delta (r.h.s.)=2\,n\,\gamma\;\int\,
             \Phi(\mid{\bf r}-{\bf y} \mid)\;
                   \left(1+u(y)\right)\,\delta u(y)\,d^3y\;.
\label{11}
\end{equation}
Using relations (\ref{10}) and (\ref{11}) and the equality
$\delta (l.h.s.)=\delta (r.h.s.)$, one can derive the following
integro--differential equation for $K_{{\bf q}}({\bf r})$:
\begin{eqnarray}
\frac{\hbar^2}{2\,m}\,\biggl(\nabla^2\,K_{{\bf q}}({\bf r})
      &-&{\displaystyle K_{{\bf q}}({\bf r})
             \over 1+u(r)}\,\nabla^2\,u(r)\biggr)=
\widetilde u(q)\,\frac{\displaystyle \exp(i
     {\bf q}{\bf r})}{(2\pi)^3}\,+\nonumber\\[5mm]
      &+&\,n\,\gamma\,\biggl(1+u(r)\biggr)
\int \Phi(\mid{\bf r}-{\bf y}\mid)\,
           \left(1+u(y)\right)K_{{\bf q}}({\bf y})\,d^3y\,.
\label{12}
\end{eqnarray}

The boundary conditions for $u(r)$ are given by
\begin{equation}
\lim\limits_{r \to \infty}\;u(r)\,=\,0\, , \quad u(0)\,<\,\infty\;.
\label{13}
\end{equation}
To know the behaviour of $K_{\bf q}({\bf r})\;,$ when
$r \rightarrow \infty$, one should keep in mind that the
boundary conditions (\ref{13}) are not changed by perturbing
the one--particle kinetic energy. This implies that for any
regular $\displaystyle \delta T(q) \ll T(q)$ we have
\begin{equation}
\delta u(r)=\int\,K_{{\bf q}}({\bf r})\,\delta T(q)\,d^3q
\rightarrow 0 \quad (r \rightarrow \infty)\;.
\label{13a}
\end{equation}
Hence, one could expect that $\displaystyle K_{{\bf q}}({\bf r})
\rightarrow 0\;(r\rightarrow \infty )\;.$ However, it is not the
case. To be convinced of this, let us investigate (\ref{6}) and
(\ref{12}) in the weak coupling regime which is introduced with
the inequalities
$$
u(r) \ll 1\;, \quad K_{{\bf q}}({\bf r}) \ll 1\,.
$$
Keeping only the leading terms in (\ref{6}) and (\ref{12}) results
in the following expressions:
\begin{equation}
\frac{\hbar^2}{m}\,\nabla^2\,u(r)=\gamma\,\Phi(r)\,+\,
2\,n\,\gamma\,\int \,\Phi(\mid{\bf r}-{\bf y}\mid)\,u(y)\,d^3y\,
\label{14}
\end{equation}
and
\begin{equation}
\frac{\hbar^2}{2\,m}\,\nabla^2\,K_{{\bf q}}({\bf r})=
\widetilde u(q)\,\frac{\displaystyle \exp(i
   {\bf q}{\bf r})}{(2\pi)^3}\,+\,n\,\gamma\,
\int \Phi(\mid{\bf r}-{\bf y}\mid)\,K_{{\bf q}}({\bf y})\,d^3y\,.
\label{15}
\end{equation}
From (\ref{14}) and (\ref{15}) it follows that the Fourier
transforms of $u(r)$ and $K_{{\bf q}}({\bf r})$ ($\widetilde u(p)$
and $\widetilde K_{{\bf q}}({\bf p})$, respectively), obey the
relations
\begin{equation}
\widetilde u(p)=-\,\frac{1}{2}\; {\displaystyle\gamma
  \widetilde \Phi(p)\over T(p)\,+\,n\,\gamma \widetilde
                                                     \Phi(p)}\;,
\label{18}
\end{equation}
\begin{equation}
\widetilde K_{{\bf q}}({\bf p})\,=\,-\, {\displaystyle
  \widetilde u(q) \over T(p)\,+\,n\,\gamma \widetilde \Phi(p)}\;
    \delta({\bf p}-{\bf q})\;.
\label{19}
\end{equation}
Remark that in the weak coupling case the value $\widetilde u(p)$
at any given $p$ depends only on the quantity of the one--particle
kinetic energy at this very $p\;.$ So, the $\delta$-function
behaviour of $K_{{\bf q}}({\bf p})$ in (\ref{19}) is a result of,
say, the "macroscopic" contribution of the $T(p)$ to $\widetilde
u(p)\;.$ With (\ref{18}) and (\ref{19}) it is not difficult to get
the expression for $\displaystyle K_{{\bf q}}({\bf r})$ in the weak
coupling regime:
\begin{equation}
K_{{\bf q}}({\bf r})\,=\,\frac{1}{16\,\pi^3}\,{\displaystyle
  \gamma\,\widetilde \Phi(q) \over \biggl(T(q)\,+\,n\,\gamma
\widetilde \Phi(q)\biggr)^2}\; \exp(i\,{\bf q}\,{\bf r})\;.
\label{20}
\end{equation}
As it is seen, for $r \rightarrow \infty$ the quantity
$K_{{\bf q}}({\bf r})$ oscillates and does not tend to zero.
However, relation (\ref{13a}) is fulfilled.

To have a reasonable idea concerning the long--range behaviour of
$K_{{\bf q}}({\bf r})$, let us note that (\ref{12}) is reduced to
(\ref{15}) not only in the weak coupling regime but in the
case $r \rightarrow \infty$ as well. This can easily be determined
via inserting $u(r)\simeq 0 \;(r \rightarrow \infty)\;$ into
(\ref{12}). Then it is reasonable to expect that the solution of
(\ref{12}) is close to that of (\ref{15}) at large distances.
Hence, investigating (\ref{12}) we may adopt the following
ansatz:
\begin{equation}
K_{{\bf q}}({\bf r}) = D_{{\bf q}}({\bf r})\,
+ C_{{\bf q}} \exp(i\,{\bf q}\,{\bf r})\;,
\label{23}
\end{equation}
where the boundary conditions
\begin{equation}
\lim\limits_{r \to \infty}\;D_{{\bf q}}({\bf r}) = 0, \quad
D_{{\bf q}}(0) < \infty
\label{24}
\end{equation}
are fulfilled for the ''short--range`` function $D_{{\bf
q}}({\bf r})$ at any ${\bf q}\,.$ Substituting (\ref{23}) into
(\ref{12}) and using (\ref{13}) and (\ref{24}), we obtain
\begin{equation}
C_{{\bf q}}=-\,\frac{1}{(2\,\pi)^3}\;{\widetilde u(q)
             \over T(q)\,+\,n\,\gamma \widetilde \Phi(q)}\;.
\label{24a}
\end{equation}
So, the ''long--range`` part in (\ref{23}) is very similar to the weak
coupling expression for $K_{{\bf q}}({\bf r})\;.$ However, $\widetilde
u(q)$ is now the solution of the exact equation (\ref{6}).

Integro--differential equations (\ref{6}) and (\ref{12}) provide
the detailed information concerning the distribution $n(q)\;.$
However, when evaluating only the condensate fraction $n_0/n$, we do
not exactly need the function $K_{\bf q}({\bf r})\;.$ It is sufficient
to work with the quantity
\begin{equation}
K(r)\,=\,\int\;K_{\bf q}({\bf r})\;d^3q\;.
\label{25}
\end{equation}
Indeed, using expressions (\ref{1}), (\ref{5}), (\ref{7}) and
the definition of $u(r)$, we derive
\begin{equation}
\frac{n_0}{n}\,=\,1-\int\limits_{0}^{1}\,d\gamma\,\int\,\Phi(r)\,
(1+u(r;\gamma,n))\,K(r;\gamma,n)\,d^3r\, ,
\label{26}
\end{equation}
where, according to the notation mentioned above, the identity $K(r)
\equiv K(r;\gamma,n)$ is meant. An equation for $K(r)$ can be
determined by integrating (\ref{12}) over ${\bf q}$, which results
in the following:
\begin{eqnarray}
\frac{\hbar^2}{2\,m}\,\biggl(\nabla^2\,K(r)
      &-&{\displaystyle K(r)
             \over 1+u(r)}\,\nabla^2\,u(r)\biggr)=
   u(r)\,+\nonumber\\[5mm]
     &+&\,n\,\gamma\,\biggl(1+u(r)\biggr)
\int \Phi(\mid{\bf r}-{\bf y}\mid)\,
           \left(1+u(y)\right)K(y)\,d^3y\,.
\label{27}
\end{eqnarray}
The boundary conditions additional to (\ref{27}) are of the form
\begin{equation}
\lim\limits_{r \to \infty}\,K(r)=0, \quad K(0) < \infty
\label{28}
\end{equation}
and follow from (\ref{23}) and (\ref{24}).

Now, let us return to the weak coupling regime to compare our results
with the data of the Bogoliubov approach. Using (\ref{5}), (\ref{7})
and (\ref{20}) we arrive at the equality
$$
n(q)=\frac{n^2}{2}\;\int\limits_{0}^{1}\;d\gamma\;
     {\displaystyle \gamma\,\widetilde \Phi^2(q)\over
           \biggl(T(q)+n\,\gamma\,\widetilde \Phi(q)\biggr)^2}
$$
which, after integration, can be written as
\begin{equation}
n(q)=f(x)=\frac{1}{2}\,\left( \ln \biggl(1+\frac{1}{x}\biggr) -
           \frac{1}{1+x}\right),
              \quad x\equiv {T(q)\over n\,\widetilde \Phi(q)}\,.
\label{29}
\end{equation}
Expression (\ref{29}) should be compared with the result of
the Bogoliubov approach that, up to the first order,
is given~\onlinecite{bog} by
\begin{equation}
n(q)=f_{Bog}(x)=\frac{1}{2}\;\left(\frac{1+x}{\sqrt{x^2+2x}}
                                               -1\right) \;.
\label{30}
\end{equation}
The values of $f(x)$ and $f_{Bog}(x)$ for $0.02 \leq x \leq 8$ are
listed in Table 1. Besides, for large $x \gg 1$ and small $x
\ll 1$ we have
$$
f(x) \simeq f_{Bog}(x) \simeq \frac{1}{4x^2} \quad (x \gg 1)\;,
$$
$$
f(x) \simeq - \frac{\ln x}{2}\;, \quad f_{Bog}(x) \simeq
                     \frac{1}{2\sqrt{2\,x}} \quad (x \ll 1)\;.
$$
So, the model considered provides quite satisfactory estimates
of $n(q)$ for $x > 0.04\;.$ However, it does not yield adequate
values of the boson distribution over momenta when $x \rightarrow
1\;.$ This feature is not unexpected because the interval of small
$x$ corresponds to the region of small $q\;.$ In other words, the
behaviour of $n(q)$ in the case $q \rightarrow 0$ is significantly
influenced by the asymptotics of $g(r)-1$ for $r \rightarrow \infty$
which is not reproduced by equation (\ref{6}) in a proper way
(see the discussion in \onlinecite{shan1,shan2}). The model of Ref.1
is able to yield good evaluations for the pair correlation function
$g(r)-1$ when $r \ll r_c$ and $r \sim r_c$ (here $r_c$ stands for
the correlation length). This means that it can provide reasonable
estimates of $n(q)$ for $q\,r_c \sim 1$ and $q\,r_c \gg 1$ and
relevant values of the macroscopic quantities such as the mean
energy or the condensate fraction $n_0/n\,.$ Let us, for example,
consider the cold many--boson system with the Coulomb interaction
in the weak coupling regime. Remark that new interest in this
system has been inspired by curious results of applying the Bose and
Bose--Fermi liquid models in the investigations of
high--temperature superconductors~\onlinecite{alex}. In the case
of charged bosons $\displaystyle\widetilde \Phi(q)=4\pi e^2/q^2$
($e$ is the boson charge) and thus $\displaystyle x=2(q/A)^4$,
where $A=(16\pi n m e^2/\hbar^2)^{1/4}$ is the inverse screening
radius. Using equalities (\ref{1}) and (\ref{29}) we obtain
\begin{equation}
\frac{n_0}{n}=1 \,-\,{\displaystyle A^3 \over (2\pi)^2\,2^{3/4}\,n}
\;\int\limits_{0}^{\infty} \left( \ln\biggl(1+\frac{1}{t^4}\biggr)
 -\frac{1}{1+t^4}\right)\,t^2\,dt.
\label{31}
\end{equation}
Integration by parts allows one to rewrite expression~(\ref{31})
in the form
\begin{equation}
\frac{n_0}{n}=1\,-\,{\displaystyle 2^{1/4}\;A^3 \over 24\,\pi^2\,n}
\;\int\limits_{0}^{\infty} \frac{t^2}{1+t^4}\,dt\, .
\label{32}
\end{equation}
Further, with the equality (see \onlinecite{grad})
\begin{equation}
\int\limits_{0}^{\infty} \frac{t^{\mu - 1}}{1+t^{\nu}}\,dt \,=\,
\frac{\pi}{\nu}\,{\displaystyle 1 \over
                               \sin\left(\mu\pi/\nu\right)}\, ,
\label{33}
\end{equation}
one can derive
\begin{equation}
\frac{n_0}{n}=1-\left( \frac{2}{3}\right)^{1/4}\,
                                \frac{r_S^{3/4}}{6} \approx
       1-0.1506\,r_S^{3/4}\; ,
\label{34}
\end{equation}
where $r_S$ stands for the Brueckner parameter given by
$$
r_S=\frac{m\,e^2}{\hbar^2} \left(\frac{3}{4\,\pi\,n}\right)^{1/3}\;.
$$
The leading term of the result for the condensate fraction in
the Bogoliubov approach~\onlinecite{fol} is of the form
\begin{equation}
\left[\frac{n_0}{n}\right]_{Bog} \approx 1 - 0.2114\,r_S^{3/4}\;.
\label{35}
\end{equation}
The reasonable character of the model considered can also be
illustrated with the calculations of the ratio of the mean
kinetic energy $\langle T \rangle$ to the mean interaction energy
$\langle U \rangle\,.$ Using (\ref{29}) and $x=2(q/A)^4$, for the
quantity $\langle T \rangle / N $ we have
\begin{eqnarray}
\frac{\langle T \rangle}{N}  &=& \frac{V}{(2\pi)^{3}}\int n(q)
                                                     T(q)\,d^3q=
\nonumber\\[3mm]
 &=& { \hbar^2\, A^5 \over (4\pi)^2 \;2^{1/4}\,m\,n}
\;\int\limits_{0}^{\infty} \left( \ln\biggl(1+\frac{1}{t^4}\biggr)
 -\frac{1}{1+t^4}\right)\,t^4\,dt.
\label{36}
\end{eqnarray}
Integrating over $t$ by parts and taking account of (\ref{33}),
one is able to find
\begin{equation}
\frac{\langle T \rangle}{N\,Ry}=\frac{1}{5}\left(
                                        \frac{3}{2}\right)^{1/4}\,
r_S^{-3/4}\approx 0.2213\,r_S^{-3/4}\; ,
\label{38}
\end{equation}
where $\displaystyle Ry=\frac{me^4}{2\,\hbar^2}\;.$ Note that when
$e$ is equal to the electron charge, we obtain $ Ry \approx 13.6\;
eV\;.$ The mean interaction energy of the system of charged bosons
calculated within the model of Ref.1 to the first order in the weak
coupling approximation, is written as
\begin{equation}
\frac{\langle U \rangle}{N\,Ry}= -\left( \frac{3}{2}\right)^{1/4}\;
r_S^{-3/4} \approx -1.107\,r_S^{-3/4}\, .
\label{39}
\end{equation}
Equality (\ref{39}) can readily be derived via the expression
$$
\langle U \rangle =\lim\limits_{\gamma \to 1}
   \int \biggl( g(r;\gamma,n)-1 \biggr) \frac{e^2}{r}\,d^3r=
\lim\limits_{\gamma \to 1}\,\frac{1}{4\pi^3}\int\,\widetilde
u(q;\gamma,n)\,\frac{4\pi e^2}{q^2}\,d^3q
$$
and with (\ref{18}) and (\ref{33}). Combining (\ref{38})
and (\ref{39}) gives
\begin{equation}
{\langle T \rangle\over\mid \langle U \rangle \mid} =
                                        \frac{1}{5}=0.2\;.
\label{40}
\end{equation}
The leading term of the mean kinetic energy in the Bogoliubov
approach can be calculated with the help of (\ref{30}) on the
analogy of deriving (\ref{36})
\begin{equation}
\frac{\hspace{0.4cm}\langle T \rangle_{Bog}}{N} ={\hbar^2\,A^5\over
(4\pi)^2\;2^{1/4}\,m\,n}
\;\int\limits_{0}^{\infty} \left( \frac{1+t^4}{\sqrt{t^8+2t^4}}
      -1\right)\,t^4\,dt.
\label{41}
\end{equation}
Numerical integration yields
$$
\int\limits_{0}^{\infty} \left( \frac{1+t^4}{\sqrt{t^8+2t^4}}
      -1\right)\,t^4\,dt \approx 0.2015,
$$
and we hence get
\begin{equation}
\frac{\hspace{0.4cm}\langle
              T \rangle_{Bog}}{N\,Ry}= 0.2008 \,r_{S}^{-3/4}\;.
\label{42}
\end{equation}
The mean energy of the charged bosons taken up to the first order
in the Bogoliubov approximation has the form~\onlinecite{fol}
\begin{equation}
\frac{\hspace{0.4cm}
           E_{Bog}}{N\,Ry} \approx - 0.8026 \,r_S^{-3/4}\,.
\label{43}
\end{equation}
Relations (\ref{42}) and (\ref{43}) lead us to the ratio
\begin{equation}
{\langle T \rangle_{Bog}\over \mid \langle U \rangle_{Bog}\mid}=
{\langle T \rangle_{Bog}\over \mid E_{Bog}-\langle T
\rangle_{Bog}\mid} \approx 0.2001
\label{44}
\end{equation}
that is in nice agreement with (\ref{40}).

Thus, considered in the weak coupling approximation, equations
(\ref{6}) and (\ref{12}) provide good estimates for the pair
correlation function and boson momentum distribution in relevant
regions of distances and momenta. This makes it possible to get
adequate evaluations of the condensate fraction and main
thermodynamic quantities of the cold many--boson system with weak
interaction between particles. Owing to the correct account of
the short--range correlations in (\ref{6}) ( see
\onlinecite{shan1,shan2}), one may expect that equations
(\ref{6}), (\ref{12}) and, hence, (\ref{27}) are able to yield
reasonable data beyond the weak coupling regime as well.

\newpage
\noindent
TABLES:\\[10mm]
\noindent
Table 1
\vskip 1cm
\begin{tabular}{|c|c|c||c|c|c|}
\hline
\hspace{0.2cm}$x$\hspace{0.5cm}& $f(x)$\hspace{0.5cm} &
                                         $f_{Bog}(x)$
                                              \hspace{0.5cm}&
\hspace{0.3cm} $x$\hspace{0.4cm} & $f(x)$\hspace{0.5cm} &
                                         $f_{Bog}(x)$
                                               \hspace{0.5cm}\\
\hline
0.02 & 1.476 & 2.037 &  1.0 & 0.097 & 0.077\\
\hline
0.04 & 1.148 & 1.320 &  1.5 & 0.056 & 0.046\\
\hline
0.06 & 0.964 & 1.008 &  2.0 & 0.036 & 0.030\\
\hline
0.08 & 0.839 & 0.824 &  4.0 & 0.012 & 0.010\\
\hline
0.10 & 0.745 & 0.700 &  6.0 & 0.006 & 0.005\\
\hline
0.50 & 0.216 & 0.171 &  8.0 & 0.003 & 0.003\\
\hline
\end{tabular}

\end{document}